\documentclass[num-refs]{nbdt-article}

\usepackage{siunitx}

\newcommand{\cm}[1]{{\color{black}#1}}

\papertype{Original Article}
\paperfield{Journal Section}

\title{The problem of perfect predictors in statistical spike train models}

\author[1]{Sahand Farhoodi}
\author[1]{Uri Eden}

\affil[1]{Department of Mathematics and Statistics, Boston University, Boston, MA, 02215, USA}

\corraddress{Sahand Farhoodi, Department of Mathematics and Statistics, Boston University, Boston, MA, 02215, USA}
\corremail{sahand@bu.edu}

\fundinginfo{NIH, NIMH, MH105174; Simons Foundation, Simons Collaboration on the Global Brain, 542971}

\runningauthor{Sahand Farhoodi et al.}

\begin{document}

\maketitle

\begin{abstract}
Generalized Linear Models (GLMs) have been used extensively in statistical models of spike train data. However, the \cm{maximum likelihood estimates of the model parameters and their uncertainty}, can \cm{be challenging to compute} in situations where response and non-response can be separated by a single predictor or a linear combination of multiple predictors. Such situations are likely to arise in many neural systems due to properties such as refractoriness and incomplete sampling of the signals that influence spiking. In this paper, we describe multiple classes of approaches to address this problem: \cm{using an optimization algorithm with a fixed iteration limit}, computing the maximum likelihood solution in the limit, Bayesian estimation, regularization, change of basis, and modifying the search parameters. We demonstrate a specific application of each of these methods to spiking data from rat somatosensory cortex and discuss the advantages and disadvantages of each. We also provide an example of a roadmap for selecting a method based on the problem's particular analysis issues and scientific goals.

\end{abstract}

\section{Introduction}
Generalized linear models (GLM) provide a powerful tool for relating observed neural spiking data to the biological, behavioral, and stimulus signals that influence them \cite{Eden_Kass_2016, Paninski_Brown_Kass_2008, Paninski_Pillow_2007}. These models express the conditional intensity of spiking, which defines the probability of observing a spike in any small time interval, in terms of a design matrix whose columns represent the signals, or covariates, of interest. GLMs possess a number of properties that make them well suited to spike train modeling \cite{Chen_2013, Paninski_Pillow_2007, Truccolo_2005, Keat_2001}. One important property is that GLMs can be formulated to guarantee that the likelihood of the spiking data is convex with respect to the model parameters, allowing for rapid computation of their maximum likelihood estimates (MLE) \cite{Paninski_Pillow_2004, Paninski_2004}. \cm{Multiple algorithms exist for maximizing the likelihood function, such as simulated annealing \cite{Bertsimas_1993, LaarhovenEmile_1987}, Fisher scoring \cite{Longford_1987}, and gradient descent methods \cite{Ruder_2016}. Because the likelihood function is convex in GLMs,} most statistical software packages implement maximum likelihood estimation for GLMs using the computationally efficient Iteratively Reweighted Least Squares algorithm (IRLS algorithm) \cite{McCullagh_1989, Green_1984, Fisher_1935}. Another useful property of GLMs is that in most cases, the parameter estimates have an asymptotically multivariate normal distribution with a covariance matrix determined by the Fisher information, which is computed as part of the IRLS algorithm \cite{McCullagh_1989, Green_1984}. This makes it easy to compute confidence intervals (CI) about individual parameters or about the firing intensity, as well as providing for simple hypothesis tests, e.g. maximum likelihood ratio tests, about whether firing rates are influenced by specific sets of covariates \cite{McCullagh_1989, Green_1984}. Accurate estimation of the Fisher information also helps determine the extent to which one signal or set of signals is confounded with another in its influence on neural spiking \cite{McCullagh_1989, Green_1984}.

For some GLMs, \cm{the algorithm to maximize the likelihood} may fail to converge because the likelihood is actually maximized in the limit as one or multiple parameters tend to $\pm \infty$. This problem is sometimes referred to as the monotone likelihood or non-existence of the MLE problem \cite{Bryson_1981}. \cm{For example, consider a point process, such that whenever a non-negative covariate $X$ is nonzero, no events occur. For a Poisson regression model relating $X$ to the rate of events, the rate should go to zero, and therefore the log rate should go to $-\infty$, whenever $X$ is nonzero. This can only be achieved by making the parameter multiplying $X$ be equal to $-\infty$.} Another perspective is that this situation arises when the response and non-response can be separated by a single predictor or a linear combination of multiple predictors. For this reason, this phenomenon is more commonly referred to as the complete separation problem \cite{Albert_1984}. This problem is well-studied in the case of logistic regression with small to medium-sized data \cite{Kolassa_1997, Lesaffre_1993, Clarkson_1991, Lesaffre_1989, Santner_1986, Albert_1984}. Heinze et. al. explore a number of potential solutions for this case including the omission of predictors that result in complete separation, fixing the parameter values for such predictors prior to fitting the model, changing the form of the model, and using an ad-hoc adjustment \cite{Heinze_2006, Heinze_2002}. They arrive at a preferred procedure using a modification proposed by Firth in order to reduce the bias of maximum likelihood estimates in GLMs \cite{Firth_1993, Firth_1992a, Firth_1992b}. 

The problem of complete separation is discussed in the statistical literature on point processes such as Cox processes as well \cite{Nagashima_2017, Heinze_Cox_2001}, \cm{and has been pointed out in the context of Poisson models for neural spiking \cite{Zhao_2010}, but remedies have not been discussed in detail in the neural modeling literature}, despite the fact that it is likely to arise in many neural systems due to properties such as refractoriness and incomplete sampling of the signals that influence spikes. In these models, separation occurs when any covariate forces the system not to spike whenever it takes on a nonzero value. Such covariates are called perfect predictors, since one can perfectly predict no spiking when they are nonzero. For example, a model for a neuron with an absolute refractory period lasting longer than 1 ms will be perfectly predicted by an indicator for whether there has been a spike in the past ms. Perfect predictors can also appear in models when nonzero values of certain predictors are infrequently sampled, even in cases where a lot of data exists. For example, a model fit for a rat's hippocampal place field that includes an indicator for being in a corner of the environment that the rat tends to avoid may suggest that this variable is a perfect predictor, even though the neuron might occasionally fire in that region given enough time. We say that perfect predictors of the kind in the first example are structural, while ones of the kind in the second example are the result of sampling.

We posit that when these issues arise in statistical neural models, researchers adopt one of a set of ad-hoc methods to avoid dealing with it, and the issues do not make the discussion of the resulting papers. However, perfect prediction can lead to a variety of issues limiting the utility of point process GLMs, which may not be alleviated by ad-hoc approaches. One important issue is that the computed Fisher information in these fit models does not accurately reflect the variance-covariance structure of the parameter estimates, \cm{which makes quantifying uncertainty challenging}. In particular, the computed Fisher information is often close to singular, and if inappropriately used to construct confidence bounds and perform hypothesis tests, typically leads to extremely large standard error estimates for the perfect predictors and inaccurate covariance between the perfect predictors and all other signals \cite{Heinze_2006, Vaeth_1985, Hauck_1977}. \cm{Another issue is that in the presence of perfect predictors, certain parameters may have estimated values that are completely determined by the properties of the estimation method. Since the true maximum likelihood estimates occur at $\pm \infty$, these parameter estimates are biased, and the amount of bias can differ substantially based on the choice of methods. Bias and variance issues become more important when the estimated parameters themselves are used to make inferences about neurobiological mechanisms of spiking. For instance, in the context of single-neuron spiking models, parameters related to the influence of past spike history have been interpreted in terms of a combination of ionic currents and a moving threshold, allowing researchers to relate inferred parameters to these biological processes \cite{Latimer_2019, Amidi_2018, Weber_2017, Latimer_2014}.} The next issue is that \cm{the algorithm used to maximize the likelihood} may not converge, and will typically only stop when some fixed iteration limit has been reached. This often involves an order of magnitude more computational time than fitting a GLM without perfect predictors. In some cases, \cm{the optimization algorithm} reaches a region of the likelihood surface where the computations become numerically unstable, and subsequent iterations can actually lead to a reduction in likelihood. 

In this paper, we explore a range of different approaches for dealing with perfect predictors in point process GLMs. These approaches fall into multiple categories: Interpreting the IRLS output despite lack of convergence, fixing the parameter estimates for perfect predictors, Bayesian estimation methods, regularization methods, re-parameterizing the model to remove perfect predictors, and modifications to the parameter search domain or stopping criteria. In the Methods section, we introduce these different approaches and explain how they deal with the problem of separation. In the Results section, we select one specific method from each category and illustrate its application to modeling spiking data from a rat cortical neuron, in vitro. Furthermore, we compare these methods and illustrate how they deal with different problems associated with complete separation. Finally, in the Discussion section, we discuss the properties of these approaches, point out some advantages and disadvantages of each method, and introduce a potential road map to select the most appropriate method to use in various situations.

\section{Methods} \label{sec:Methods}
\subsection{Problem Formulation}
Let $X$ be the design matrix with $p$ predictors of neural spiking (columns) and $n$ observations (rows) and $Y = (y_1, \ldots, y_n)^{T}$ be the response vector of spike counts where the superscript $T$ denotes the transpose of a matrix. We use $x_{ij}$, $x_{i\cdot}$ and $x_{{\cdot}j}$ respectively to denote the $ij$-th element, $i$-th row and $j$-th column of matrix $X$. Let $\beta = (\beta_0, \beta_1, \ldots, \beta_{p})^T$ be the parameter vector, $\theta = X \beta$, and $\lambda = \mathbb{E}(Y|X)$ be the conditional expectation of $Y$ given $X$. Let $g(\cdot)$ be the link function that relates the conditional expectation of $Y$ to the linear combination of predictors, i.e. $g(\lambda) = \theta$. Here, we work with the canonical link function for Poisson point process which is the log function \cite{McCullagh_1989}. The Poisson process GLM is described by
\begin{subequations}\label{eq:PP_model}
\begin{eqnarray}
    &y_i|x_{i{\cdot}} \sim \textrm{Poiss}(\lambda_i) \label{eq:PP_model1} \\[10pt]
    &\lambda_i = \textrm{exp} \Big(\beta_0 + \sum\limits_{j=1}^{p} x_{i j} \beta_j \Big) \label{eq:PP_model2}.
\end{eqnarray}
\end{subequations}
where $\lambda_i$ is the firing rate for the $i$-th observation and $1 \leq i \leq n$. 
\cm{Perfect prediction presents challenges for any algorithm that is focused on maximizing the likelihood when its surface is nearly flat over a large subset of the parameter space. This includes approaches such as simulated annealing \cite{Bertsimas_1993, LaarhovenEmile_1987}, Fisher scoring \cite{Longford_1987}, and gradient descent methods \cite{Ruder_2016}. Because of the convex likelihood surface in GLMs,} the IRLS algorithm is widely used to find the maximum likelihood solution in this case. This algorithm starts with an initial vector $\beta^{(0)}$ and then computes $\beta^{(s)}$ recursively using the update equation 
\begin{eqnarray}
    \beta^{(s+1)} = \beta^{(s)} + I\big(\beta^{(s)}\big)^{-1}U\big(\beta^{(s)}\big) \label{eq:IRLS_update}
\end{eqnarray}
until convergence where $U(\beta)$ and $I(\beta)$ respectively denote the score function and the Fisher information computed for $\beta$. For the model described in Eq. \ref{eq:PP_model}, the log-likelihood function $l(\beta)$, the score function $U(\beta)$, and the Fisher information $I(\beta)$ are defined as
\begin{subequations}\label{eq:like_score_Fisher}
\begin{eqnarray}
    &l(\beta) = \sum\limits_{i=1}^{n} \big( y_{i}\textrm{log}(\lambda_i) - \lambda_i \big) \label{eq:likelihood}\\[10pt]
    &U(\beta) = \frac{\partial l(\beta)}{\partial \beta} = X^{T}(Y - \lambda) \label{eq:score_function}\\[10pt]
    &I(\beta) = -\frac{\partial^2 l(\beta)}{\partial \beta^2} = X^{T}WX \label{eq:Fisher_information}
\end{eqnarray}
\end{subequations}
where $W$ is the matrix of weights defined by $W^{-1} = (\frac{d \theta}{d \lambda})^2 V$ and $V$ is the variance function evaluated at $\lambda$ \cite{McCullagh_1989}. For our choice of link function $g(\cdot)$, $W$ will be a diagonal matrix with $i$-th diagonal element equal to $\lambda_i$.

The $j$-th column of $X$ is a perfect predictor if having a nonzero value at this column leads to zero spike count, i.e. for any $i \in \{1, \ldots, n$\}, $x_{ij} \neq 0$ implies $y_i = 0$. Additionally, a set of columns, $j_1, \ldots, j_m$, generate a perfect predictor if some linear combination of these are a perfect predictor, i.e. there exists a set of weights $a_1, \ldots, a_m$, such that whenever $a_1 x_{ij_1} + \cdots + a_m x_{ij_m} \neq 0$ then $y_i = 0$. \cm{We let $S$ be the set of indices corresponding to columns that generate perfect predictors, and define the set of perfect parameters as $\{\beta_j : j \in S \}$. The maximum likelihood estimators for these parameters diverge to $\pm \infty$. We say that the $i$-th row is perfectly predicted if there is $j \in S$ such that $x_{ij} \neq 0$.}

A naive way to deal with perfect predictors is to omit them from the model by removing all \cm{perfectly predicted rows and all columns corresponding to perfect parameters} from $X$. However, perfect predictors are extremely informative and removing them can harm the model substantially in terms of goodness-of-fit and statistical power. Putting this naive approach aside, in what follows we briefly explore different families of approaches that can be used to deal with perfect predictors. 

\subsection{\cm{Fixed Iteration Limit}}
The first approach for dealing with perfect predictors is to use \cm{an optimization algorithm such as the IRLS algorithm} to estimate the MLE without adjusting the model or estimation procedure. As pointed out before, in the presence of perfect predictors the actual maximum likelihood solution is achieved only when perfect parameters are equal to $\pm \infty$, and therefore the parameter estimates will depend on when the algorithm terminates, typically after a fixed iteration limit is reached.

\subsection{Maximum Likelihood Limit}
Eq. \ref{eq:PP_model2} can also be rewritten as
\begin{eqnarray} \label{eq:rem_PP_model}
    \lambda_i = \textrm{exp} \Big(\beta_0 + \sum\limits_{j \notin S} x_{i j} \beta_j + \sum\limits_{j \in S} | x_{i j} | \beta_j \Big).
\end{eqnarray}
For perfect parameters ($j \in S$) taking the absolute value of $x_{ij}$ only affects the sign of $\beta_j$ and doesn't change the predicted values of our model. Thus, the maximum likelihood solution is achieved when for any $j \in S$, $\beta_j$ is equal to its limiting value $- \infty$. \cm{Therefore, one approach is to set perfect parameters equal to their limiting value manually, and then use an algorithm that maximizes the likelihood, e.g. the IRLS algorithm, to estimate non-perfect parameters, after omitting perfect predicting columns and perfectly predicted rows from $X$ and $Y$.} The final model then includes both the estimated parameters when perfect predictors are removed, and the perfect parameters set to $- \infty$. In practice, a perfect parameter cannot be set numerically equal to -$\infty$, so often a very large negative number is used instead.

\subsection{Bayesian Estimation}
The next approach to deal with the problem of separation is using Bayesian GLM \cm{\cite{Gerwinn_2010, Penny_2005}}. In this approach, we treat the parameters $\beta$ as random variables, assign them prior distributions $\pi$, and compute and maximize their posterior distributions given the observed data. For the case that $\pi$ is a zero-mean \cm{normal} distribution with covariance matrix $\Sigma$, the posterior log-likelihood function, the posterior score function, and the posterior Fisher information are respectively given by
\begin{subequations}\label{eq:Bayesian_like_score_Fisher}
\begin{eqnarray}
    &l_{b}(\beta) = \sum\limits_{i=1}^{n} \big( y_{i}\textrm{log}(\lambda_i) - \lambda_i \big) - \frac{1}{2}\beta^T\Sigma^{-1}\beta \label{eq:Bayesian_likelihood}\\[10pt]
    &U_{b}(\beta) = X^{T}(Y - \lambda) - \Sigma^{-1}\beta \label{eq:Bayesian_score_function}\\[10pt]
    &I_b(\beta) = X^{T}WX + \Sigma^{-1} \label{eq:Bayesin_Fisher_information}
\end{eqnarray}
\end{subequations}
where the subscript $b$ stands for Bayesian. In order to use the IRLS algorithm to fit Bayesian GLM, these equations can be used in Eq. \ref{eq:IRLS_update} to maximize $l_b(\beta)$. An appropriate choice of $\Sigma$ can solve the perfect predictor problem in a couple of ways: the diagonal terms of $\Sigma$ prevent any individual parameters from diverging to $\pm \infty$, and the off-diagonal terms of $\Sigma$ can be used to impose smoothness between sets of parameters, so that sampling issues are less likely to cause perfect predictors.

\subsection{Regularization Methods}
Another approach that is commonly used to address the problem of separation is regularization. In this approach, instead of maximizing the log-likelihood function, a weighted average of the log-likelihood function and a penalty function, $p(\beta)$, is maximized. This penalty function is intended to prevent parameters from diverging by penalizing larger values of the parameters. Lasso and Ridge penalty functions are two of the most used \cite{Hastie_2001}. A penalty function that has been used specifically to deal with the problem of separation in logistic regression is Firth's penalty function \cite{Heinze_2002}. In fitting a regularized GLM, the regularized log-likelihood function, the regularized score function, and the regularized Fisher information are computed according to
\begin{subequations}\label{eq:regularization_like_score_Fisher}
\begin{eqnarray}
    &l_{r}(\beta) = (1-\Lambda) l(\beta) - \Lambda p(\beta) \label{eq:regularized_likelihood}\\[10pt]
    & U_{r}(\beta) = (1-\Lambda)X^{T}(Y-\lambda) - 2\Lambda \frac{d p(\beta)}{d \beta} \label{eq:regularized_score_function}\\[10pt]
    &I_{r}(\beta) = (1-\Lambda)X^{T}WX - 2\Lambda \frac{d^2 p(\beta)}{d \beta^2} \label{eq:regularized_Fisher_information}
\end{eqnarray}
\end{subequations}
where the subscript $r$ stands for regularization and $\Lambda$ is a variable determining the amount of penalization. \cm{To extend the IRLS procedure,} these quantities can be used directly in Eq. \ref{eq:IRLS_update} to maximize the regularized log-likelihood function. \cm{It is worth mentioning that some Bayesian approaches provide equivalent estimates as regularization methods with specific penalty functions \cite{Park_2008, Yuan_2005, Tibshirani_1996}. For instance, the Bayesian GLM described in Eq. \ref{eq:Bayesian_like_score_Fisher} can be framed as a regularization method with the penalty function $p(\beta)=\frac{1}{2}\beta^T \Sigma^{-1} \beta$.}

\subsection{Change of Basis}
The next family of methods focuses on reparameterizing the model in order to impose smoothness between specific sets of parameters. This method can be used when we believe that the receptive field has some smoothness properties that are not explicitly captured by the model structure. For example, it is common to use estimated spike rates in nonoverlapping spatial bins to model place fields, despite the fact that this does not capture the fact that rates in adjacent bins tend to be close. Similarly, models of refractoriness or bursting often focus on autocovariance in lagged bins, ignoring the tendency of neurons to have smooth history dependence structure. One way of imposing smoothness on predictors is using basis functions such as smoothing splines \cite{Lin_2006, Cook_1981} or raised cosines \cite{Pillow_2008}. Employing this approach Eq. \ref{eq:PP_model2} is replaced by
\begin{eqnarray}\label{eq:spline_PP_model}
    \lambda_i = \textrm{exp} \Big(\tilde{\beta}_0 + \sum\limits_{k=1}^{q} g_k(x_{i.}) \tilde{\beta}_k) \Big)
\end{eqnarray}
where $g_k$ is the $k$-th basis function and $q$ denotes the number of basis functions. \cm{Any standard optimization algorithm such as the IRLS algorithm described in Eq. \ref{eq:IRLS_update} can be used to fit this model.} This approach is capable of alleviating issues related to both structural and sampling perfect predictors, by introducing a new smoother predictor, $g_k(x_{i.})$, that integrates information over a range of the receptive field that is better sampled.

\subsection{Modified Search Criteria}
\cm{Another approach, sometimes referred to as constrained optimization, is to modify the search or stopping criteria for the estimation algorithm to prevent parameters from diverging \cm{\cite{Uryasev_2013, Rossi_2008, Box_1965}}.} One explicit way of doing so is to restrict the search space of the \cm{optimization algorithm} and identify optimal parameters within this restricted space. Another way of preventing the estimated parameters from diverging is to \cm{require parameter updates to improve the likelihood by at least a threshold value. For example, if we use the IRLS algorithm to maximize the likelihood, we can} update $\beta_j$ in Eq. $\ref{eq:IRLS_update}$ only if $U(\beta_j)$ is greater than a selected threshold. This prevents the IRLS algorithm from making adjustments to $\beta$ when the log-likelihood function is nearly flat, which is a characteristic of the likelihood in the presence of perfect predictors.

\section{Results } \label{sec:Results}

\subsection{Data Set and the Specific Estimation Algorithms}
In this section, we pick one specific method from each family of approaches introduced in the Methods section, and compare them for the problem of fitting spiking data collected from rat cortical neurons\cm{\footnote{\cm{The results of this section are generated using MATLAB, and the code is accessible in https://github.com/SahandFarhoodi/Perfect\textunderscore Predictors.}}}. This data has been previously used for challenge A of the Quantitative Single-Neuron Modeling Competition (2009) developed by researchers at Ecole Polytechnique Fédérale de Lausanne (EPFL) and is accessible on CRCNS.org \footnote{See http://crcns.org/data-sets/challenges/ch-epfl-2009}. It contains spiking activity from the somatosensory cortex of a Wistar rat in response to somatic current stimulation. This dataset contains 13 repetitions of a 60-second stimulation protocol, where for the first 39 seconds both the injected current waveform in pA and voltage responses are provided \cm{for bins of length 1 ms} \cite{Gerstner_2009, CRCNS}. We only use the first 39 seconds of these repetitions to construct our training set. Two repetitions are used to fit the model and the other repetitions are used for resampling to compute cross-validation statistics and to examine whether perfect predictors arise due to subsampling. Different pairs of repetitions are used to fit each model multiple times (10 times in total) and \cm{the mean and standard deviation of each performance measure is reported in Table \ref{tab:summary_table}.}

We denote the firing rate of the neuron at time $t$ by $\lambda(t)$, the spiking count in the interval $(t,t+\Delta t)$ by $\Delta N_{t}$, the firing history of the neuron up to time $t$ by $H_t$, and let $p$ be a fixed value representing the length of history which we believe can influence the current spike intensity. We divide the whole range of current stimulus levels into $q$ disjoint intervals, and let $I_j(t)$ be an indicator function that takes value 1 if the input current at time $t$ falls into the $j$-th interval. We relate the firing rate at time $t$ to the past history of spike counts and the input current using the following point process GLM:
\begin{eqnarray}
    \lambda(t \mid H_t) = \textrm{exp} \Big( \beta_0 + \sum_{j=1}^{p} \beta_j \Delta N_{t-j} + \sum_{j=1}^{q} \beta_{p+j} I_j(t) \Big). \label{eq:applied_model}
\end{eqnarray}
where $p$ and $q$ are set to 200 ms and 6 respectively. In general, statistical model identification procedures, such as the step-wise regression, can be used to select the order of a model. Here, our purpose is not to identify the most parsimonious model, but to introduce a model that can capture specific features of the data, and compare multiple fitting procedures in their handling of perfect predictors.

When we fit this model for a single neuron using the IRLS algorithm directly, we obtain a fit with 31 perfect predictors. 20 of these perfect predictors are associated with the parameters that capture history dependence (first sum of Eq. \ref{eq:applied_model}) occurring at small lags due to the refractory period of the neuron. These likely reflect structural perfect predictors, in that we would not expect to see any spikes occur immediately after another, even if we collected much more data. There are also two perfect predictors corresponding to the two lowest current indicator functions, $I_1(t)$ and $I_2(t)$. It is not immediately clear if these are structural perfect predictors or sampling predictors (i.e. whether the neuron has a non-negligible probability of firing when the stimulus current is this low). In this case, when we increase the data size (by using the extra trials in the training set to fit the model), these perfect predictors remain so, suggesting they may be structural as well. The remaining perfect predictors occurred either in the history component of the model at specific lags that had neighbors that were not perfect predictors, or in the stimulus response component for stimulus levels that we expect could drive some amount of neural spiking. In order to analyze if these perfect predictors are structural or due to sampling, we expand the training data by including additional trials and fit the same model described in Eq. \ref{eq:applied_model}. We observed that all these perfect predictors vanish and hence categorize them as sampling perfect predictors, since they only existed in the first place because of limited sampling in the training data. Without using any of the methods described in the Methods section, all the perfect predictors in this example diverged to large negative numbers, limited only due to the finite iteration limit of the IRLS algorithm. 

In order to compare the general approaches defined in the Methods section, we selected one specific method from each family that, in our estimation, is the most natural option or is most likely to be used by neural data analysts when encountering the problem of separation. However, we focus only on those advantages and disadvantages of each method that can be generalized to all methods of the family from which it was selected. Below, we specify the method selected from each family of approaches.

\begin{description}
\item[\cm{Fixed Iteration Limit.}] We use the IRLS algorithm which is widely used for fitting point process GLMs. We set the iteration bound equal to 100, and run the IRLS algorithm without adjusting the model or estimation procedure. We call this specific method "Standard IRLS" throughout the rest of this paper.
\item[Maximum Likelihood Limit.] In this method, \cm{the perfectly predicted rows and the columns associated with perfect parameters} are removed from $X$ and $Y$ prior to fitting the model. The final model includes the estimated parameters when perfect predictors are removed, combined with perfect parameters set to $-\infty$.
\item[Bayesian Estimation.] There are two types of parameters in the model described by Eq. \ref{eq:applied_model}, the ones corresponding history spiking counts ($\beta_{1:p}$) and the ones corresponding input current ($\beta_{p+1:p+q}$). We consider separate prior distributions for these two groups, as we observed no correlation between their parameters. Each one of these prior distributions is a zero-mean multivariate Normal distribution with a covariance matrix $\Sigma$ with elements $\Sigma_{ij} = c^{\mid i-j \mid}$ where $c \in [0, 1]$. This choice of $\Sigma$ defines a temporal correlation on parameters that fall off geometrically as the distance between them increases, and therefore can deal with the problem of separation by imposing smoothness between nearby estimated parameters. We selected the parameter $c=0.9$ using a leave-one-out cross-validation method. We note that more advanced methods for estimating this parameter are possible (e.g. treating it as a hyperparameter and building a hierarchical Bayesian model to estimate it).
\item[Regularization.] We used ridge regression, which uses an $L_2$ penalty term, to fit the model parameters because of its common usage in the neural data analysis literature. Therefore, the penalty function in Eq. \ref{eq:regularization_like_score_Fisher} will be $p(\beta) = \beta^T \beta$. The choice of $\Lambda$ can affect the outcome of the regularization model significantly and thus, we use cross-validation to determine it. The value that gave the best model fit was $\Lambda = 0.1$. This specific method is called "Ridge GLM" throughout the rest of this paper.
\item[Change of Basis.] We reparameterized the model using a cardinal spline basis expansion \cite{Lin_2006, Cook_1981} in place of the spike count and indicator functions in Eq. \ref{eq:applied_model}. As shown in Appendix 1, this basis expansion method is capable of eliminating perfect predictors if the knots are chosen such that there is at least one non-perfect predictor between every two successive knots. However, by choosing knots too far from each other, a good amount of information in the data will be lost due to excessive imposed smoothness. This makes it challenging to determine the ideal number and location of the knots. One complicated but proven to work option is to use Bayesian curve-fitting with free-knot spline algorithms \cite{Dimatteo_2001}. Another simpler option is letting knots be distributed regularly and then using cross-validation only to select the number of knots. In this case study, we used the latter approach, where we pick the numbers of knots that gives the best goodness-of-fit. This specific method is called "Cubic Smoothing Spline" throughout the rest of this paper. 
\item[Modified Search Criterion.] As an example of a modified search criterion approach, we restrict the search space of the IRLS algorithm to $R = \{\beta : \sum_{j=1}^{p+q} \beta_i^2 \leq r \}$. We set $r = (p+q)d^2$, where $p+q$ is the total number of parameters (excluding the intercept) and $d = -5$ is the threshold for identifying if a parameter is perfect or not based on the output of the Standard IRLS method. We call this method "Bounded Search" in the rest of this paper.
\end{description}

\subsection{Comparison of Specified Methods}
\begin{figure}[t]
\centerline{\includegraphics[width=\linewidth]{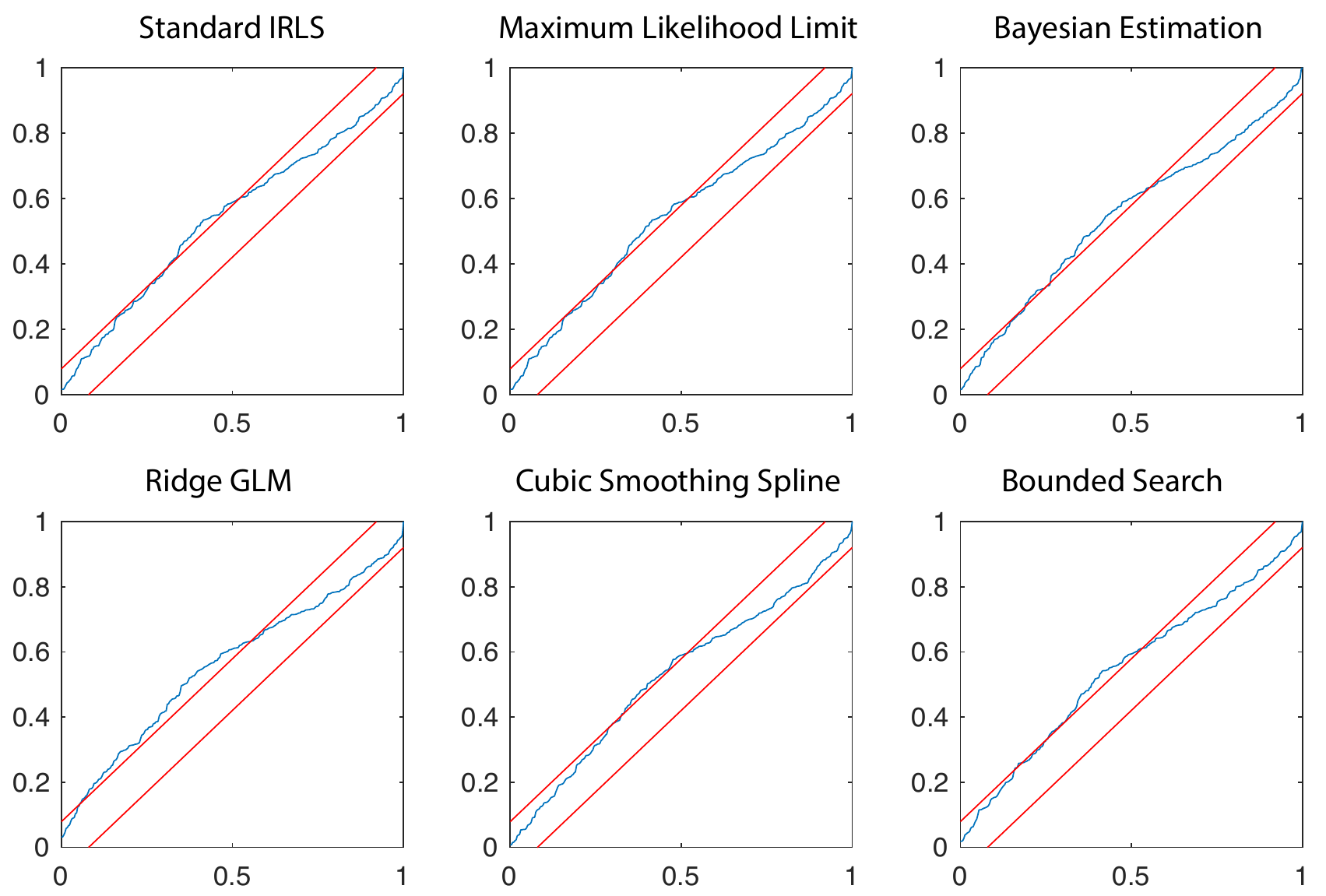}}
\caption{KS plots for six different fitting approaches. The spike times were transformed via the time-rescaling theorem \cite{Brown_2002, Papangelou_1972} based on each fitted model. The blue line is the KS plot and the red lines represent global 95\% confidence bounds for a well fit model. The Cubic Smoothing Spline method shows the smallest deviation outside of the bounds and the Ridge GLM has the largest, suggesting the best and the worst performance respectively.}
\label{fig:KS_test}
\end{figure}
First, we compare the different methods based on their goodness-of-fit. One approach for investigating goodness-of-fit is to examine Kolmogorov-Smirnov (KS) plots comparing the empirical and theoretical distributions of transformed spike times based on the times-rescaling theorem \cite{Brown_2002, Papangelou_1972}. Fig. 1 depicts these plots for different methods we specified in the previous section. The blue line represents the KS plot and the red lines represent global 95\% bounds for a well-fit model. The larger the deviation outside of these bounds, the poorer the model fit to the data. The Cubic Smoothing Spline provides the best fit and Ridge GLM provides the poorest fit to the data from this neuron in terms of the KS analysis. The rest of the methods are fairly similar based on the KS plots. 

Another approach for comparing goodness-of-fit between models is to examine the proportion of deviance in a null model that is explained by each, which is expressed by
\begin{eqnarray} \label{eq:Dev_ratio}
    \cm{R} = \frac{D_{N} - D_{M}}{D_{N}}. 
\end{eqnarray}
Here, $D_M$ and $D_N$ denote the deviance \cm{\cite{McCullagh_1989}} of the proposed model and the null model respectively, \cm{and the null model is a point process with constant intensity. For a Poisson GLM with the canonical log link function, the deviance is given by
\begin{eqnarray}
    D = 2\sum_{i=1}^n y_i \log \left(\frac{\hat{\lambda}_i}{y_i}\right) + (y_i - \hat{\lambda}_i)
\end{eqnarray}
where $\hat{\lambda}$ is the estimated intensity from the model.} The value of $R$ shows the ratio of the deviance explained by the proposed model that the null model fails to explain. This measure does not attempt to account for over-fitting, and tends to be higher for larger models. Therefore, we also consider a cross-validated version of $R$, denoted by $R_{CV}$, which can directly be used to assess the prediction power of the fitted models. A statistical comparison between different models based on their deviances requires taking into account the effective degrees of freedom (d.o.f.) of each model as well. Among various measures introduced for the effective d.o.f., we use the approach that to our knowledge is the most prevalent. This approach defines the effective d.o.f to be the trace of the hat matrix $H$, which is satisfying $HY = \hat{Y}$ where $\hat{Y}$ denotes the estimate of the response vector \cite{Hastie_2001}. In order to make the comparison easier, we divide the effective d.o.f. for each method by that for Standard IRLS, and call it effective d.o.f. ratio. Additionally, we compare the different methods based on the computational resources they require. To do so, we compute the relative run time and the relative peak memory of each method with respect to standard IRLS on the same system. All models have been fit 10 times, each using a different pair of trials from the training set, and mean \cm{and standard deviation} values are computed.

Table \ref{tab:summary_table} summarizes the comparisons between different methods, in terms of goodness-of-fit and consumed computational resources. \cm{The values in the table represent means over the 10 simulations and the values in parentheses represent standard deviations.} Of all the methods, the Standard IRLS and Maximum Likelihood Limit methods have the largest $R$ values for the data used to fit the model, and the Cubic Smoothing Spline has the smallest. However, these methods do not generalize as well to other datasets, as they have the smallest $R_{CV}$ values. In fact, $R_{CV}$ for the standard IRLS approach is on average negative, indicating a poorer fit than a null, homogeneous Poisson model of the data (see Eq. \ref{eq:Dev_ratio}). This suggests that the effect of leaving the perfect predictors in the model is to overfit the observed data, with large magnitude perfect parameter values that make a few observed non-spike intervals slightly more likely, but lead to extremely poor fits when future datasets include spikes for which the fitted model predicts no chance of spiking. On the other hand, all of the other methods impose some kind of constraint on the estimated parameters, e.g. via a prior distribution in Bayesian Estimation or via smoothing in the Cubic Smoothing Spline approach, which prevents the sampling perfect parameters from diverging. We observe that Cubic Smoothing Splines and Bayesian Estimation, which have noticeable smaller effective d.o.f ratios, have better predictive power (larger $R_{CV}$) than other methods. The Ridge GLM is the slowest method (largest relative run time), which is due to its exhaustive search to find an optimal value for the tuning parameter $\Lambda$. \cm{The Bayesian Estimation approach also spends a noticeable amount of time on finding the optimal value for the tuning parameter $c$ through cross validation, which makes it the next slowest approach.} Cubic Smoothing Splines, despite the fact that it has the smallest effective d.o.f ratio, shows the largest relative peak memory after Standard IRLS. This is because this method needs to construct and store a new design matrix before running the IRLS algorithm. \cm{We note that the variability of the $R_{CV}$ and relative run time suggest that for this example differences in these measures are not likely due to chance. While this is true for this specific example, these trends may or may not hold for other analysis problems and model structures. A more systematic analysis would be necessary to determine statistical significance between methods for any particular inference problem.}

\begin{table}[bt]
\caption{Summary of goodness-of-fit measures and computational resource consumption for different estimation methods. All models have been fit 10 times, each time using a different pair of repetitions as the training set, and mean values are computed, \cm{along with standard deviation values shown inside parenthesis.}}
\begin{threeparttable}
\begin{tabular}{lcccccc}
\headrow
\thead{Method} & \shortstack{$R$} & \shortstack{$R_{CV}$} & \thead{\shortstack{Number of \\ Parameters}} & \thead{\shortstack{Effective \\ d.o.f. Ratio}} & \thead{\shortstack{Relative \\ Run Time}} & \thead{\shortstack{Relative \\ Peak Memory}}\\
Standard IRLS & 0.3404 (.005) & $<$ 0  (.007) & 206 & 1 & 1 (0) & 1 \\
Maximum Likelihood Limit & 0.3404 (.005) & 0.1479 (.007) & 175 & 0.83 & 0.07 (.05) & 0.07 \\
Bayesian Estimation & 0.3227 (.003) & 0.2827 (.02) & 206 & 0.45 & 0.16 (.06) & 0.01\\
Ridge GLM & 0.3236 (.002) & 0.2319 (.02) & 206 & 0.79 & 1.95 (.15) & 0.05\\
Cubic Smoothing Spline & 0.2612 (.002) & 0.2418 (.02) & 40 & 0.18 & 0.09 (.06) & 0.18\\
Bounded Search & 0.3397 (.005) & 0.2124 (.009) & 206 & 0.86 & 0.15 (.08) & 0.01\\
\hline
\end{tabular}
\end{threeparttable}
\label{tab:summary_table}
\end{table}

Another issue that accompanies the complete separation or perfect prediction problem is that the observed Fisher information at the maximum likelihood solution typically does not accurately reflect the variance-covariance structure of the parameter estimates. In particular, the observed Fisher information is often close to singular, and if inappropriately used to construct confidence bounds and perform hypothesis tests, typically leads to extremely large standard error estimates for the perfect predictors and inaccurate covariance between the perfect predictors and all other signals \cite{Heinze_2006, Vaeth_1985, Hauck_1977}. Below, we examine how different each of the methods deals with this problem.

\begin{figure}[bt]
\centerline{\includegraphics[width=\linewidth]{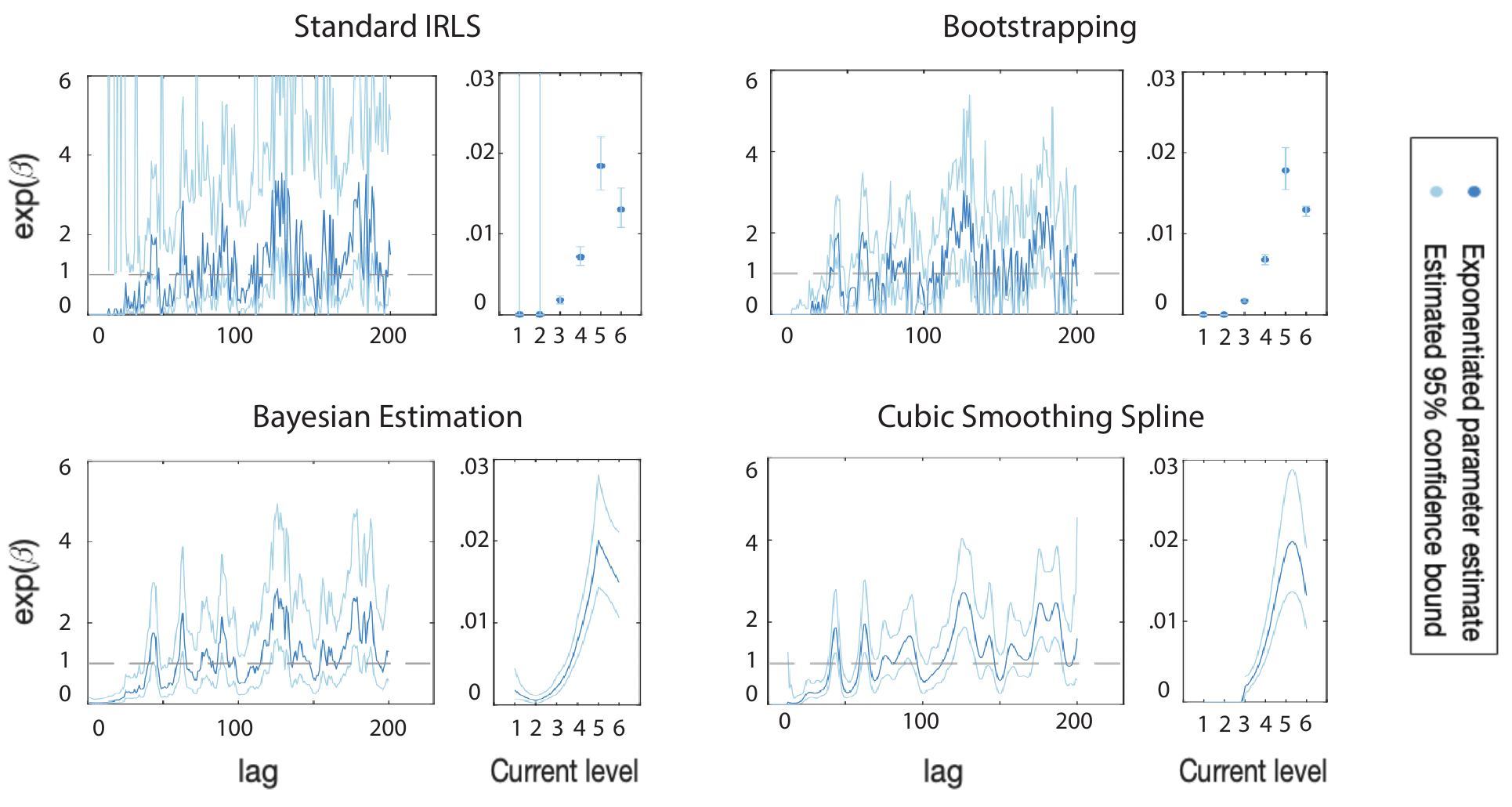}}
\caption{Estimated parameter values, $\hat{\beta}$, and estimated confidence intervals obtained by four different approaches: Standard IRLS, Bootstrapping, Bayesian Estimation, and Cubic Smoothing Splines. For each approach, the estimated influence of past spiking is shown in the left panel and the estimated influence of the input current is shown in the narrower right panel. Thick blue lines or dots represent the exponentiated parameter estimates and thin blue lines and bars represent the estimated 95\% confidence bounds.}
\label{fig:CI}
\end{figure}

Bootstrapping represents one alternative to the Fisher information for computing standard error estimates for perfect predictors for the Standard IRLS estimator. In this approach, random subsets of the original data are used to fit the model multiple times, and then the empirical distribution of estimated values is used to compute standard error estimates for all parameters. Bootstrapping is far more computationally expensive than the other proposed methods, but results in more robust standard error estimates for IRLS, and therefore is considered as our benchmark here. The estimated values, $\hat{\beta}$, along with their confidence intervals using Standard IRLS, Bootstrapping, Bayesian Estimation, and the Cubic Smoothing Spline approaches are shown in Fig. \ref{fig:CI}. As expected, the bootstrap approach results in no standard error estimates for structural perfect parameters, as they always diverge to $-\infty$. The bootstrap intervals are tighter than those obtained by the Fisher Information for IRLS for nearly all parameters, suggesting that the existence of perfect predictors can influence uncertainty estimates in the IRLS procedure even for non-perfect predictor parameters. Both Bayesian Estimation and Cubic Smoothing Spline methods result in relatively narrow confidence intervals which provide smoother estimates than bootstrapping IRLS. We observed that Ridge GLM and Bounded Search obtain standard error estimates that are artificially larger than what was obtained by Bootstrapping, especially for structural perfect parameters. The Maximum Likelihood Limit approach doesn't provide an estimation procedure for standard error of perfect parameters, as they are completely removed from the model prior to fitting.

\begin{figure}[bt]
\centerline{\includegraphics[width=1\linewidth]{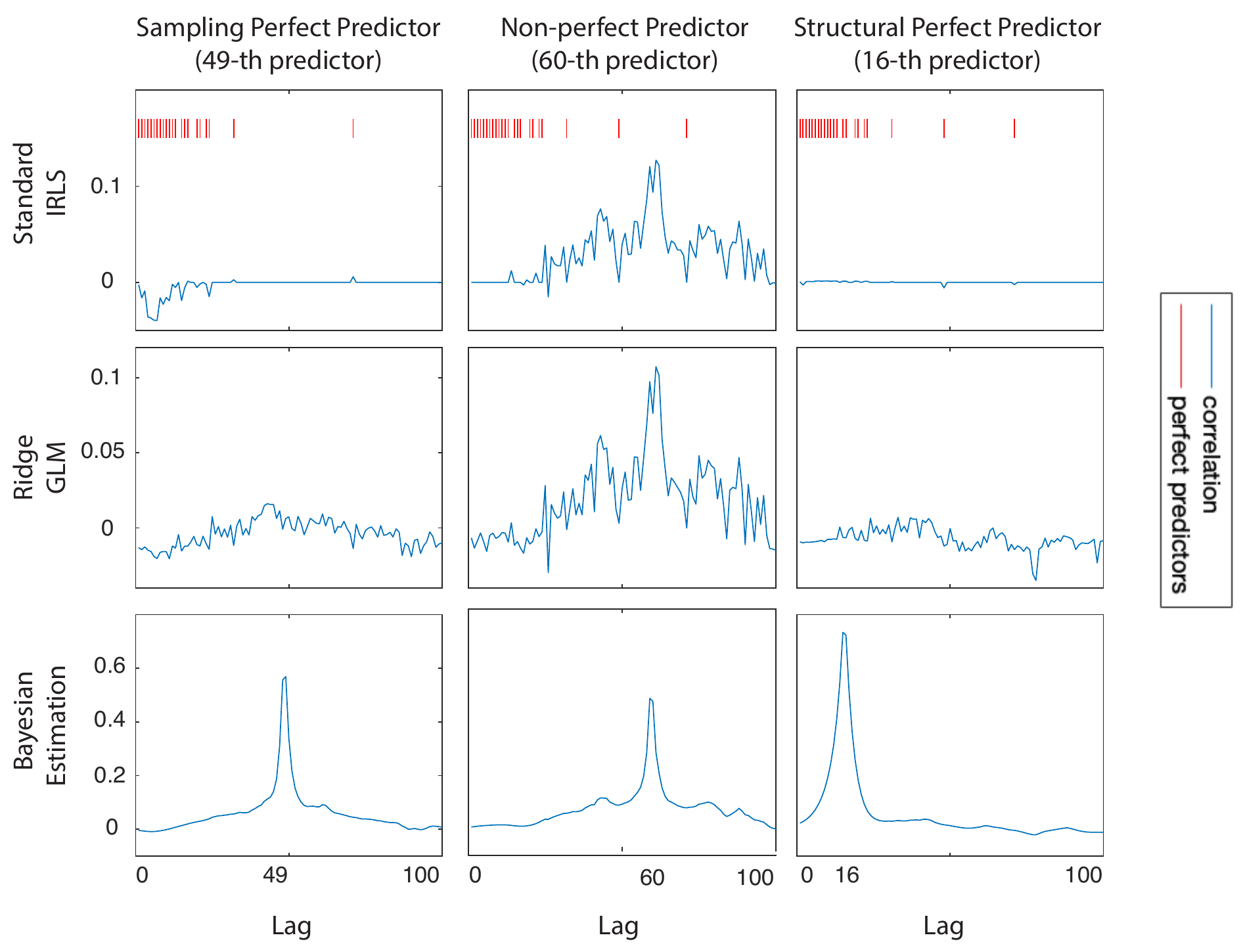}}
\caption{Estimated correlation between a fixed predictor and all other predictors associated with the past history of spike counts at lags 1-100ms. This is shown for a sampling perfect predictor (left column), a non-perfect predictor (middle column), and a putative structural perfect predictor (right column), when Standard IRLS (first row), Ridge GLM (second row), and Bayesian Estimation (third row) were used. The red ticks in the top panel represent the lags associated with perfect predictors.}
\label{fig:pred_corr}
\end{figure}

In Fig. \ref{fig:pred_corr} in order to examine how different methods estimate the correlation between parameters, we concentrate on the first 100 predictors, which are associated with the influence of past spike counts at lags of 1-100 ms, and illustrate the estimated correlation between one fixed predictor and all other predictors. For the fixed predictor, we consider a sampling perfect predictor (left column), a non-perfect predictor (middle column), and a putative structural perfect predictor associated with the neuron's refractory period (right column). We observe that for non-perfect predictors, all methods perform more or less similarly, with Bayesian Estimation providing a smoother version of the correlation structure estimated by the Ridge GLM and Standard IRLS methods (middle column). In this case, we can see a local correlation structure for parameters around the fixed predictor (60-th predictor), which is expected; the effect of firing activity at lag $i$ and $i+1$ is likely to be correlated. For the sampling perfect predictor (left column), we see that Standard IRLS results in zero correlation for all non-perfect predictors. However, we know that this predictor is perfect only due to sampling, and hence we expect to see a temporal structure similar to what we observed between two non-perfect predictors in the middle column. Ridge GLM shows this local correlation structure weakly, but with substantially reduced effect size and statistical power. On the other hand, the correlation estimated by Bayesian GLM is much more aligned with what we expect to see, and clearly shows a temporal correlation structure similar to that between non-perfect predictors. For the putative structural perfect parameter, Standard IRLS again estimates all the correlations to non-perfect predictors to be zero. In this case, this might be the preferred inference; if it is impossible for the neuron to fire during its refractory period, this should be true regardless of the influence of a spike beyond the refractory period. In this case, Ridge GLM also gives very small correlations, but Bayesian GLM still estimates a smooth correlation structure, consistent with the prior distribution. Note that we could use our knowledge of the refractory period of the neuron to adjust the prior to prevent this correlation. Though not shown in this figure, the Cubic Smoothing Spline estimates correlations very similar to the Bayesian Estimation method, and the Bounded Search estimates are very close to those obtained from Ridge GLM. The Maximum Likelihood Limit approach, again, does not provide correlation estimates between any perfect predictor and any other predictor, as it removes all perfect predictors from the model prior to fitting.

\section{Discussion}
In this paper, we investigated the problem of complete separation in GLMs, which leads to the failure of the IRLS algorithm to converge when a single predictor or a linear combination of multiple predictors completely separate a response from a non-response. This occurs frequently in point process GLMs due first, to structural properties of neurons and their receptive fields that prevent a response, e. g. refractoriness, and second, incomplete sampling of the signals that influence spiking leading to no observed neural responses. This phenomenon can have a substantial impact on modeling and statistical inference from neural data. We broadly presented various classes of approaches to deal with this problem (section \ref{sec:Methods}), and illustrated how they compare in practice (section \ref{sec:Results}). Here, we further discuss some of the advantages and disadvantages that are associated with each family of approaches.

\begin{description}
\item[Standard IRLS] \leavevmode   
\begin{description}
  \item[Advantages.] Already implemented in many statistical software packages. Describes the existing data set optimally. Inference for non-perfect parameters is typically reliable.
  \item[Disadvantages.] Very slow; continues to run until iteration limit reached. Fitted models generalize poorly to other datasets. Poor estimates of variance-covariance structure for sampling perfect parameters make inferences about these parameters challenging.
\end{description}

\item[Maximum Likelihood Limit] \leavevmode
\begin{description}
  \item[Advantages.] Fast. Describes the existing data optimally. Inference for non-perfect parameters is typically reliable.
  \item[Disadvantages.] Fitted models generalize poorly to other datasets. Not typically implemented in statistical software packages. No explicit computation of the Fisher information for estimating the variance-covariance structure of the perfect parameters.
\end{description}

\item[Bayesian Estimation] \leavevmode
\begin{description}
  \item[Advantages.] Provides increased statistical power and prediction accuracy when appropriate priors are available. Allows for precise estimation of confidence bounds and covariance between parameters. Prior can be chosen to distinguish between covariates that may lead to structural or sampling perfect predictors.
  \item[Disadvantages.] Requires methods for selecting appropriate priors. Results may be sensitive to the choice of prior.
\end{description}

\item[Regularization Methods] \leavevmode
\begin{description}
  \item[Advantages.] Already implemented in many software packages. Controls perfect parameters with minimal impact on non-perfect parameters.
  \item[Disadvantages.] Sensitive to tuning parameter, which typically involves an exhaustive search to find an optimal value. This makes this method relatively slow. Can result in diminished statistical power (compared to Bayesian Estimation and Change of Basis) since it treats structural and sampling perfect predictors identically.
\end{description}

\item[Change of Basis] \leavevmode
\begin{description}
  \item[Advantages.] Leads to more parsimonious models that take advantage of known structure in receptive field models. With a judicious choice of model structure, this can lead to a substantial increase in statistical power and model interpretability and reductions in computational time. Prior knowledge can be used to select a model structure that distinguishes structural and sampling perfect predictors.
  \item[Disadvantages.] Selecting an appropriate model structure can be challenging. Some reparameterizations may not solve the perfect prediction problem. The results may be quite sensitive to the choice of model structure. Inferences may require inversion of transformations associated with reparameterization. 
  
\end{description}

\item[Modifying Search Criterion] \leavevmode
\begin{description}
  \item[Advantages.] Has minimal impact on non-perfect parameters.
  \item[Disadvantages.] Results may be sensitive to the choice of criterion. It can result in diminished statistical power (compared to Bayesian Estimation and Change of Basis) since it treats structural and sampling perfect predictors identically.
\end{description}

\end{description}

\begin{figure}[!htb]
\centerline{\includegraphics[width=1\linewidth]{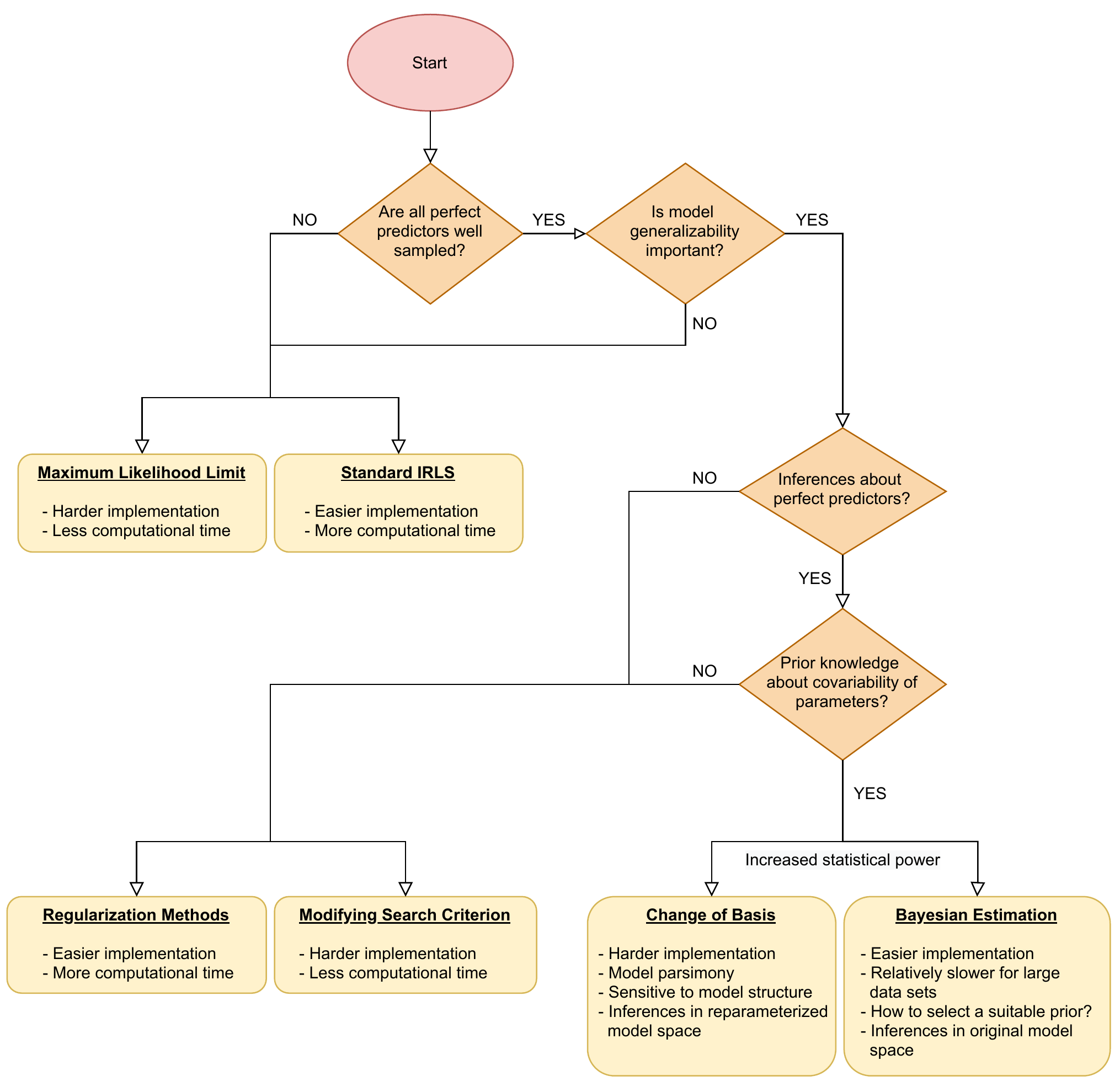}}
\caption{Roadmap for selecting an approach for dealing with perfect prediction, based on the features and goals of the particular analysis question. Each path ends with two approaches to consider with properties that are well-aligned to the resulting analysis issues. The decision between these two can be made according to the pros and cons given for each method (yellow boxes).}
\label{fig:roadmap}
\end{figure}

Knowing the potential advantages and disadvantages of each of these approaches can help modelers select one, or a subset of methods to use to explore their data.
We can also use these to develop a sequence of questions that modelers might ask to help guide them to a decision about which methods might work best for a particular analysis problem. We present an example of such a modeling `roadmap' in Fig. \ref{fig:roadmap}, in flowchart form. \cm{Note that this represents just one potential approach to selecting a method; depending on what particular set of questions might arise in the modeling process, there may be many other roadmaps, and some of these alternative roadmaps may lead to a combination of described methods, e. g. Bayesian estimation with change of basis}. In the road map depicted in Fig. \ref{fig:roadmap}, the central questions focus on how well the signals that influence spiking are sampled, whether the fitted model needs to generalize to other datasets, whether the perfect predictors relate to the scientific questions to be addressed, and whether there is prior knowledge about receptive field structure to constrain the model. Each path ends with two approaches to consider with properties that are well-aligned to the resulting analysis issues. The decision between these two can be made according to the pros and cons given for each method (yellow boxes). For example, if we had a model for a place cell using indicator functions over space, the Bayesian Estimation approach might be most useful if we have a sensible prior and the scientific questions are focused on inferences about individual parameters that represent coding over a small subregion of the place field. If sensible priors are hard to come by, or if the scientific questions are more focused on more global properties of the place field (e.g. its center and width) than a smoothed estimate based on a change to a spline basis might be more appropriate.

\cm{In some studies the objective is not only to estimate the coding properties of neurons, but also to infer physiological features of the underlying neural system. 
For instance, in the context of single-neuron spiking models, parameters related to the influence of past spike history have been interpreted in terms of a combination of ionic currents and a moving threshold, allowing researchers to relate inferred parameters to these biological processes \cite{Latimer_2019, Amidi_2018, Weber_2017, Latimer_2014}. In this case, model parameter estimates of $-\infty$ may lead to substantively different inferences than ones with large negative values, even if both lead to nearly identical predictions of the spike intensity at all times. Therefore, it is important for modelers to understand how bias related to the choice of approach can affect such physiological inferences. For this example, the maximum likelihood limit approach might be used and parameter estimates of $-\infty$ might suggest that physiological inferences are not reliable, or the Bayesian approach might be used, with priors based on physiological knowledge related to currents and threshold dynamics.}

Perfect predictors can arise in neural data due to many different factors, such as structural properties of the receptive field, lack of data for some regions of the receptive field, or even features of the receptive field that cannot be sampled under some experimental conditions. In this paper, we focused on point process GLMs, but complete separation can also happen in other domains of neuroscience. In particular, GLMs are used for Binomial and multinomial data which occur quite often when one looks at behavioral measures in neuroscience experiments. The ideas presented in this paper are not only limited to point process GLMs, and can directly be applied to other studies that involve other kinds of GLMs. Neuroscience experiments going forward are likely to yield larger data sets with many more neurons, and potentially with receptive field structures that involve many predictors simultaneously. As a result, models needed to explain such large and complex data sets are more likely to encounter the problem of complete separation. We anticipate that the types of methods and decision procedures discussed in this paper are going to become more and more important as the experiments and data sets evolve to capture more complex structures of the stimuli.

\section*{acknowledgements}

The work for this project was supported by the Simons Collaboration on the Global Brain, 542971 and the National Institute of Mental Health, MH105174. \cm{We also thank CRCNS data sharing website, and Thomas Berger, Richard Naud, Henry Markram, and Wulfram Gerstner, for collecting the data for challenge A of the 2009 quantitative single-neuron modeling competition (the data used in this paper), and publishing it publicly on CNCRS.org.}

\printendnotes

\bibliography{refs}

\section{Appendix}
\subsection{Cardinal Spline Reparameterization}
In this section, we explore a specific example of basis expansion methods, cardinal spline reparameterization, and explain in detail how it can solve the problem of perfect predictors. This basis expansion method is perfectly suitable for cases where predictors are a set of sequential indicator functions, e.g. in a history-dependent model where the $k$-th predictor is the spiking counts at lag $k$. Following the notations introduced in Eq. \ref{eq:spline_PP_model}, the $k$-th basis function is given by
\begin{eqnarray} \label{eq:spline_g_func}
    g_k(x_{j.}) = \sum_{i=1}^p x_{ji}s_{ik}
\end{eqnarray}
where $s_{ik}$ is the $ik$-th element of matrix $S$ that is constructed as follows. First, an increasing sequence of length $q-4$ from interval $(1, p)$ is augmented by 1, $p$, one value less than 1, and one value greater than $p$ to yield the sequence $\zeta_1 < \zeta_2 < \cdots < \zeta_q$ which is called the set of control points or knots. Then, in order to construct the $j$-th row of $S$, index $i$ is selected such that $\zeta_i \leq j < \zeta_{i+1}$ and $\alpha = \frac{j - \zeta_i}{\zeta_{i+1} - \zeta_{i}}$ is computed. All elements of $S_{j.}$ are set to zero except the four consecutive elements $\{s_{j, i-1}, s_{j,i}, s_{j, i+1}, s_{j, i+2}\}$ that are given by
\begin{align} \label{eq:cubic_spline_matrix}
\begin{bmatrix}
s_{j,i-1} & s_{j,i} & s_{j,i+1} & s_{j,i+2}
\end{bmatrix}
= \begin{bmatrix}
1 & \alpha & \alpha^2 & \alpha^3
\end{bmatrix}
\begin{bmatrix}
    -t & 2-t & t-2 & t\\
    2t & t-3 & 3-2t & -t\\
    -t & 0 & t & 0\\
    0 & 1 & 0 & 0\\
\end{bmatrix}
\end{align}
where $0 \leq t \leq 1$ is called the tension parameter and must be picked prior to constructing $S$. In order to find the relation between original parameters ($\beta_i$ in Eq. \ref{eq:PP_model2}) and parameters after changing basis ($\tilde{\beta}_i$ in Eq. \ref{eq:spline_PP_model}), we start by simplifying the right hand side of Eq. \ref{eq:spline_PP_model}.
\begin{align}
    \lambda_j &= exp\Big(\tilde{\beta}_0 + \sum_{k=1}^{q} g_k(x_{j.}) \tilde{\beta}_k)\Big) = exp\Big(\tilde{\beta}_0 + \sum_{k=1}^{q} \tilde{\beta}_k (\sum_{i=1}^p x_{ji} s_{ik})\Big) \nonumber\\
    &= exp\Big(\tilde{\beta}_0 + \sum_{i=1}^{p} x_{ji} (\sum_{k=1}^q s_{ik} \tilde{\beta}_k)\Big). \label{eq:beta_tilde}
\end{align}
Comparing Eq. \ref{eq:PP_model2} and Eq. \ref{eq:beta_tilde} and noting that the second sum in Eq. \ref{eq:beta_tilde} is the $i$-th element of vector $S \beta$ reveals the relation between $\beta$ and $\tilde{\beta}$:
\begin{align}
    \beta_0 = \tilde{\beta_0}, \quad \beta_{1:p} = S \tilde{\beta}_{1:q} \label{eq:spline_transform_back} 
\end{align}
Based on this relation, any new predictor is a linear combination of original predictors, and as long as some non-perfect predictor appears with nonzero weights in this combination, the new predictor will not be perfect. This can be assured to happen by choosing the knots such that there exist at least one non-perfect predictor between every two successive knots.

\end{document}